\documentclass[10pt,twocolumn,pra,
 superscriptaddress, longbibliography, 
 aps,showkeys,floatfix]{revtex4-2}
\usepackage{setspace}
\usepackage{enumitem}
\usepackage{ dsfont }
\usepackage[pdftex]{hyperref,color,graphicx}
\usepackage{amsfonts,amssymb,amsmath}
\usepackage[separate-uncertainty=false]{siunitx}
\usepackage[english]{babel}
\usepackage[utf8]{inputenc}
\usepackage[T1]{fontenc}
\usepackage[normalem]{ulem}
\usepackage[toc,page]{appendix}
\usepackage[space]{grffile}
\usepackage{placeins}
\usepackage{latexsym}
\usepackage{textcomp}
\usepackage{longtable}
\usepackage{multirow,booktabs}
\usepackage{url}
\hypersetup{colorlinks=false,pdfborder={0 0 0}}
\usepackage{dcolumn}
\usepackage{bm}
\usepackage{dsfont}
\usepackage{upgreek}
\usepackage{cancel}
\usepackage{braket}
\usepackage{xcolor}

\usepackage[normalem]{ulem}
\newcommand{\edit}[1]{\textcolor{black}{#1}}

\newcommand{\beq}{\begin{equation}}
\newcommand{\eeq}{\end{equation}}

\begin{document}

\title{Reversing adiabatic state preparation \edit{in few-level quantum systems}}

\author{Leonardo Romanato}
\affiliation{Department of Mathematical, Physical and Computer Sciences, University of Parma, Parco Area delle Scienze 7/A, 43124, Parma}

\author{N. Eshaqi-Sani}
\affiliation{Department of Mathematical, Physical and Computer Sciences, University of Parma, Parco Area delle Scienze 7/A, 43124, Parma}

\author{Luca Lepori}
%\email{luca.lepori@unipr.it}
\affiliation{Department of Mathematical, Physical and Computer Sciences, University of Parma, Parco Area delle Scienze 7/A, 43124, Parma}
\affiliation{Instituto di Fisica Nucleare (INFN), Sezione di Milano Bicocca, Gruppo Collegato di Parma, Parco Area delle Scienze 7/A, 43124 Parma}

\author{Teodora Kirova}
%\email{teodora.kirova@lu.lv}
\affiliation{Institute of Atomic Physics and Spectroscopy, Faculty of Science and Technology, University of Latvia, Jelgavas street 3, Riga, LV-1004, Latvia}

\author{Ennio Arimondo}
%\email{ennio.arimondo@unipi.it}
\affiliation{Dipartimento di Fisica, Universit\`a di Pisa, Largo Pontecorvo 3, 56127 Pisa, Italy}
\affiliation{Istituto Nazionale di Ottica - Consiglio Nazionale delle Ricerche,  Universit\`a di Pisa, Largo Pontecorvo 3, 56127 Pisa, Italy}

\author{Sandro Wimberger}
\email{sandromarcel.wimberger@unipr.it}
\affiliation{Department of Mathematical, Physical and Computer Sciences, University of Parma, Parco Area delle Scienze 7/A, 43124, Parma}
\affiliation{Instituto di Fisica Nucleare (INFN), Sezione di Milano Bicocca, Gruppo Collegato di Parma, Parco Area delle Scienze 7/A, 43124 Parma}
%NON RIESCO A FARE IL GRAFICO
\begin{abstract}
We present a detailed study of an adiabatic state preparation in an effective three-level quantum system. States can be prepared with high speed and fidelity by adding a counterdiabatic quantum control protocol. As a second step, we invert the preparation protocol to get back to the initial state. This describes an overall cyclic evolution in state space.
Using counterdiabatic terms, the resulting composed fast evolution can be repeated many times. We then analyze the control of Berry's phase along the adiabatic cyclic path and show that Berry's phase can act as a sensitive detector of non-perfect state transfer.
\end{abstract}
\date{\today}
\maketitle

\section{Introduction}
\label{sec-intro}

The question of whether a system is reversible is of fundamental interest in statistical physics \cite{LS1999}. Non-reciprocal devices, which break the time-invariance symmetry, have become fundamental in photonic, atomic as well as molecular systems.  In the conventional way of realizing non-reciprocity a magnetic field is added or magnetic materials are used. Over the past few years, non-reciprocity based on different approaches such as nonlinear systems, synthetic magnetism systems, non-Hermitian systems, and time-modulated systems have gained significant theoretical and experimental attention \cite{Koch2010, SounasAlu2017, Wu2022}.  Closed-contour driving represents an alternative approach to breaking time reversal for few-level quantum systems, see e.g. \cite{Koch2010, RoushanMartinis2017, Barfuss2018, Barfuss2019, Paraoanu2019, TaoYu2021, DV2022}.  Within this context, the dynamical aspect of the reversibility question is relevant to test the quality and repeatability of quantum gates \cite{Delvecchio2022-1}. 
 
In this paper, we study protocols to adiabatically prepare a target state and check, in particular, how such protocols behave when returning to the initial state, again under adiabatic conditions.  We correct the time evolution by transitionless or counterdiabatic (CD) driving \cite{Demirplak2003, Demirplak2005, Demirplak2008, Berry2009, Unanyan1997}. The latter approach allows one, as a specific shortcut to adiabaticity \cite{DelCampo2013, Sels2017, Odelin2019}, to reach high fidelities in the state preparation in much shorter times.  Therefore, in comparison with the traditional adiabatic evolution, the CD drive gives significant advantages for quantum technologies, where speed and fidelity (which also implies coherence) are crucial, see e.g. \cite{Barfuss2019, Daley2023, Xu2024, Lubasch2024,gentini2024}. 

After the adiabatic target-state preparation, we modify the sweep function, such as to return to the initial state, altogether following a closed contour in parameter space. Inverting the adiabatic evolution, the fidelity of the evolved state with the target one will depend on how adiabatically the evolution has been performed. Transitions to other instantaneous eigenstates of the time-dependent Hamiltonian lead to state diffusion and hence to loss when inverting the evolution. Since CD control induces adiabaticity, we reach higher fidelities for shorter sweeps, allowing for fast protocols. This then permits repeated operations on our system in reasonable total time, which is relevant for practical applications.

 The cycles in the space state resulting from the described two-steps adiabatic protocol are characterized via the associated Berry phase \cite{Berry1984, Berry1987, Wilczek1989, Zak1989, bernevig2013}, whose stability against protocol perturbations is quantitatively discussed. Our results show that we can control both the population and the Berry phase in our system.

\edit{Here we focus on a specific system introduced in ref. \cite{Delvecchio2022} that explored control schemes based on the Autler-Townes effect \cite{ATor}. The goal was} a large transfer efficiency to a final triplet state, in a system of interacting (via the spin-orbit coupling) singlet and triplet levels. Our work further develops this study for both the forward and backward evolution in the preparation of a molecular target state. The quantum control schemes presented here are easily adapted to similar setups of few-level quantum systems, which are otherwise difficult to reach, or for which reversibility is an important issue. \edit{Alternatively, other optimal control methods \cite{Koch2022} may be applied to few-level systems, as in ref. \cite{Luis2025}, which compares our counterdiabatic with a brute-force numerical control approach.}

The paper is organized as follows: Sec. \ref{sec-model} introduces our three-level problem and the considered sweep functions. Numerical results, without and with CD acceleration, are presented in Sec. \ref{sec:res}. At the end of this section, we introduce Berry's phase and a variation of our cyclic protocol that allows us to control it.
Sec. \ref{sec:sum} summarizes the paper.

\begin{figure}[tb]
    \centering
    \includegraphics[width=1\linewidth]{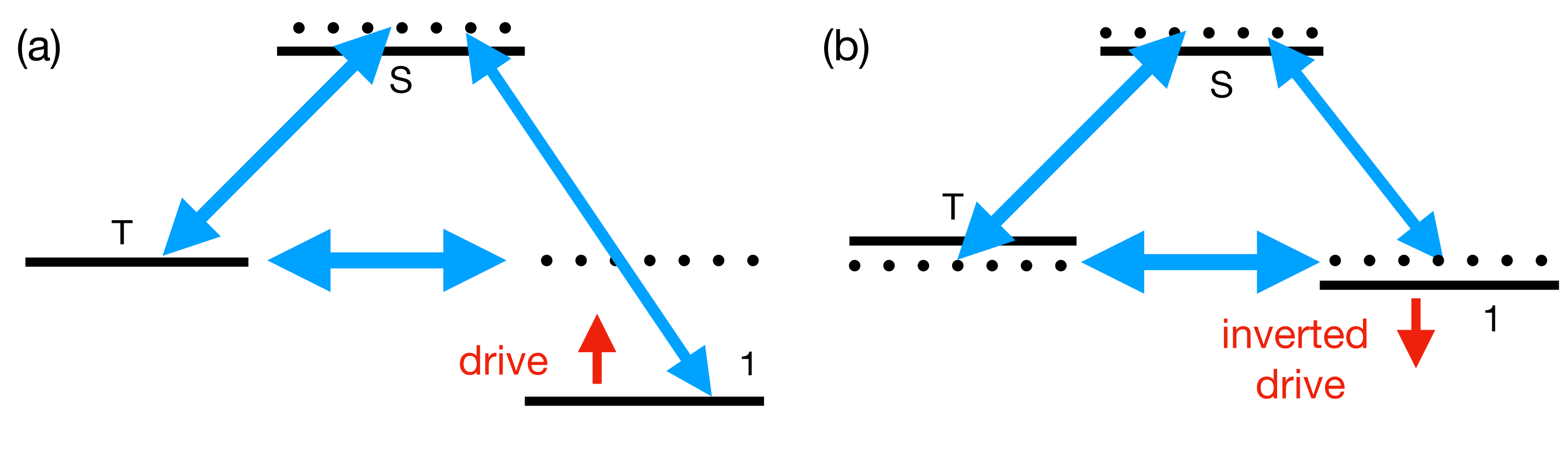}
    \caption{Sketch of the considered three state system in the bare, undriven basis. The evolution under driving occurs from
    the situation (a) to (b). The level $\ket{1}$ moves towards level $\ket{T}$, see the red arrow in (a), and both states form a tiny avoided crossing in (b), visible also in the inset of Fig. \ref{fig:2}. In a second step of the protocol the inversion of this drive brings the system back to its initial state $\ket{1}$, see the red arrow in (b).}
    \label{fig:0}
\end{figure}

\section{Model system and protocol}
\label{sec-model}

\begin{figure}[tb]
    \centering
    \includegraphics[width=1\linewidth]{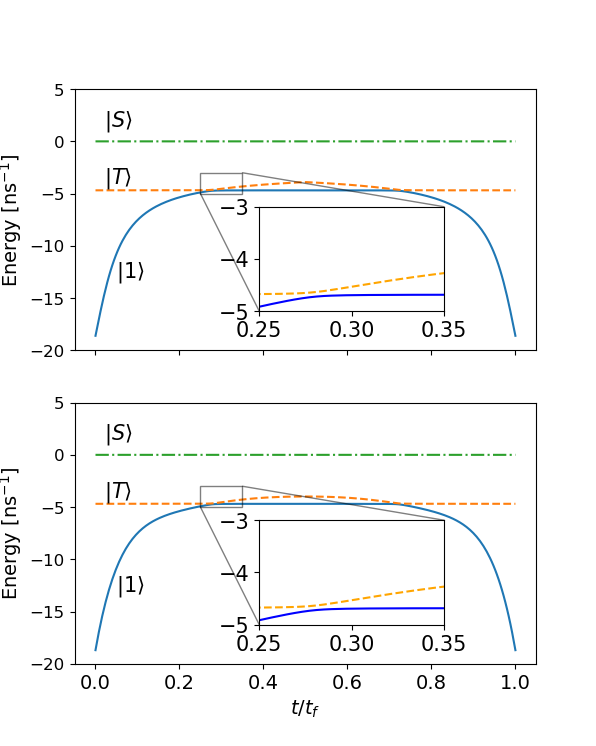}
    \caption{Instantaneous eigenvalues of the Hamiltonian from Eq. \eqref{eq:Ham-1}, for both the considered sweep functions in Eqs. \eqref{eq:atan} and \eqref{eq:fit}: $f(\tau)$ (a) and $p(\tau)$ (b). The insets show a zoom around  the first avoided level crossing we encounter during the evolution. Here no difference is visible between the two sweeps. The adopted protocol, discussed in Sec. II, aims to follow adiabatically the blue line along the time evolution. Please keep in mind that the labels of the bare basis $\{1,T,S\}$ only make sense far way from the avoided crossings, i.e., at $\tau=t/t_f\approx 0$ and $\tau\approx 1$.}
    \label{fig:1}
\end{figure}

\edit{What we show in the following applies to any three-level quantum system in which two states interact strongly locally in time and the third state is only weakly coupled to the other two levels. Motivated by previous work \cite{Delvecchio2022}, we nevertheless would like to point out a specific system for which we can use definite parameter values, and hence make specific predictions for its evolution. Hence, following Ref. \cite{Delvecchio2022},} we start from the four-level system introduced in \cite{Kirova2005}, relevant for controlling certain molecular states using the Autler-Townes mechanism \cite{ATor}. Details on the specific implementations are found in \cite{Ahmed2011, AhmedLyyra:2013, AhmedLyyra:2014}. \edit{For completeness, a short review on the model is given in appendix \ref{app:A}}. Under the condition of large detuning of the fourth state ($\vert \delta_{c}\vert\gg \Omega_{c}$ in the notation of \cite{Delvecchio2022}), this state can be adiabatically eliminated. Therefore, the system can then be effectively considered as a three-level system, with the following Hamiltonian in the basis of the three  remaining states $\{\vert 1 \rangle,\vert S \rangle,\vert T \rangle\}$:
\begin{equation}
\hat{H}_0(t)=\begin{pmatrix}
\epsilon_{1}(t) & \Omega_{1S} & \Omega_{1T}
\\
\Omega_{1S}^* & \epsilon_{S} & \Omega_{ST}
\\
\Omega_{1T}^* & \Omega_{ST}^* & \epsilon_{T}.
\end{pmatrix}.
\label{eq:Ham-1}
\end{equation}
Assuming $\hbar=1$, time and energy scales are expressed in ns and ns$^{-1}$, respectively, in what follows, just as in \cite{Ahmed2011, Delvecchio2022}.
We assume that $\epsilon_{1}(t)$ can be controlled in time. All the other matrix elements are constant, and their expressions in the notation of \cite{Delvecchio2022} are given in appendix \ref{app:A-2}. Figure \ref{fig:0} sketches the setup and the evolution 
from essentially uncoupled states $\ket{1}$ and $\ket{T}$  in (a) to a strong coupling between them at the avoided level crossing induced by the sweep of $\epsilon_{1}(t)$ in (b). 
The system is prepared in the initial state $\ket{1}$ and the aim is to achieve an efficient adiabatic population transfer from the state $\ket{1}$ to the triplet state $\ket{T}$ without populating the singlet state $\ket{S}$. In this paper, we extend this control idea to the case of an adiabatic evolution back to the initial state $\ket{1}$. For this task, we have to invert the applied sweep functions from \cite{Delvecchio2022}, once the triplet state is obtained. Let us introduce the rescaled time $\tau=\frac{t}{t_{f}}$ ($\tau\in [0,1]$), where $t_{f}$ is the final time of our protocol. We use the arctan function from \cite{Delvecchio2022} as a starting point. This function proved to give good results in \cite{Delvecchio2022} and is still reasonably simple. Our modified sweep is given piecewise as
\begin{equation}
    \label{eq:atan}
        \epsilon_{1}(\tau) = f(\tau) \equiv
        \begin{cases} 
            a \arctan(b \, \tau) - c & 0 \leq \tau < 0.5 \, , \\
             a \arctan \big(b \, (1 - \tau)\big) - c & 0.5 \leq \tau \leq1 \, ,
        \end{cases}
\end{equation}
for real parameters $a$, $b$, and $c$. The transfer to the state $\ket{T}$ occurs in the interval $\tau=0, \dots, 0.5$, and back to the initial state in $\tau=0.5, \dots, 1$. A slightly softer evolution is obtained by a polynomial fit of degree eight of the function $f(\tau)$, that reads
\begin{align}
\label{eq:fit}
p(\tau) &= a_8\tau^8 + a_7\tau^7 + a_6\tau^6 + a_5\tau^5 + a_4\tau^4 + a_3\tau^3 + a_2\tau^2 + \nonumber \\
& + a_1\tau + a_0,
\end{align}
where the coefficients depend on the parameters $a,b$ and $c$. 
These sweeps are realized by ramping the detuning of the pump laser in the setup discussed in \cite{Delvecchio2022, Ahmed2011, AhmedLyyra:2013, AhmedLyyra:2014}. They adiabatically lift the energy of state $\ket{1}$ to interact with the energy level $\ket{T}$ around $\tau=0.25$ and vice versa around $\tau=0.75$. This interaction leads to symmetrical population exchange around $\tau=0.5$. The polynomial sweep has the advantage that the discontinuity of the derivative at the inversion point $\tau=0.5$ is cured. This allows for faster adiabatic evolution and also no issues when computing the counterdiabatic Hamiltonian, presented in the next section \ref{sec:CD}. Figure \ref{fig:1} shows the instantaneous energy levels as a function of $\tau$, illustrating the structure that we have described above. Along the adiabatic evolution, the blue line in Fig. \ref{fig:1} is always followed. The avoided crossing shown in the insets corresponds to the situation sketched in Fig. \ref{fig:0}(b).

\edit{We note in passing that our problem is different from other three-level counterdiabatic control schemes, such as stimulated Raman adiabatic passage, see e.g. \cite{Petiziol2020, GiannelliArimondo2014, Chen2016}, since in our case effectively only two levels participate in the evolution. This will be demonstrated in Sec. \ref{sec-phase} below.}

\section{Results}
\label{sec:res}

In the following, we use the same experimentally relevant parameters as in ref. \cite{Delvecchio2022}. Those translate into the couplings  $\Omega_{1S}=0.11196,  \, \Omega_{ST}=0.01158, \, \Omega_{1T}=-0.0432$, and the diagonal elements $\epsilon_{S}=0.00447$ and $\epsilon_{T}=-4.74001$, all in units of inverse ns. For the sweep in Eq. \eqref{eq:atan}, we then obtain the optimized parameters $a=10$ns$^{-1}$, $b=20$ and $c=18.6$ns$^{-1}$, giving the coefficients of the polynomial which are listed in appendix \ref{app:B}.

\subsection{State preparation revisited}
\label{sec-sub-fid}
 
\begin{figure}[tb]
   \centering
   \includegraphics[width=\linewidth]{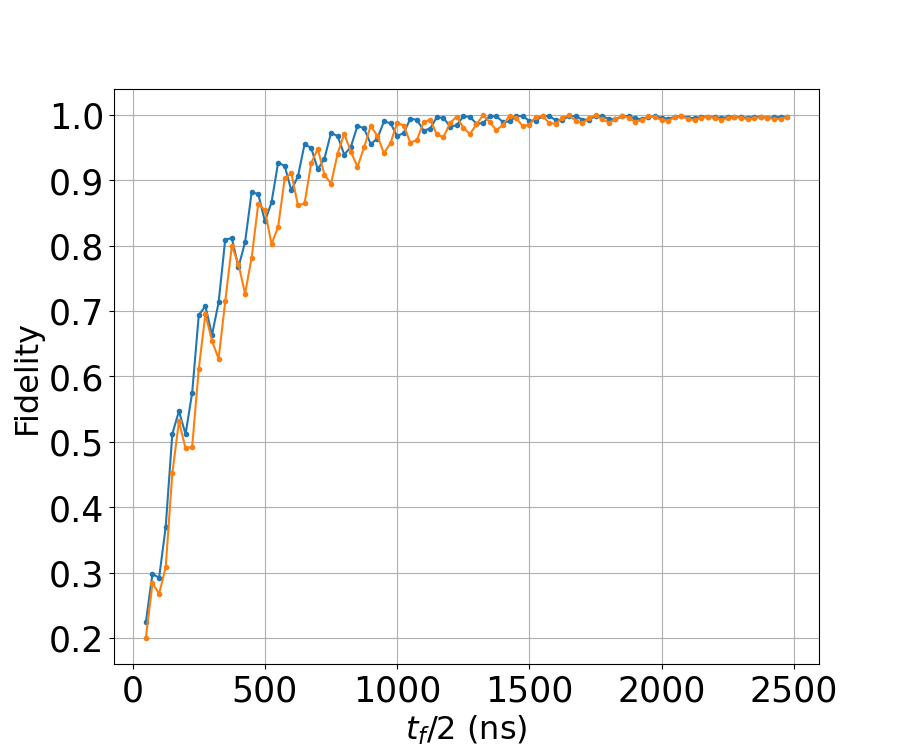}
   \caption{Overlap $F(t=t_f/2)$ with the target state $\ket{T}$, for different $t_f$ and sweeps $f(\tau)$ (blue line) and $p(\tau)$ (orange line).}
   \label{fig:2}
\end{figure}

The performance of our protocol in transferring the population from the initial state $\ket{1}$ to the target state $\ket{T}$ can be measured by the time-dependent fidelity defined as 
\begin{equation}
\label{eq:fidelity}
F(t)=\vert\langle T\vert \psi(t)\rangle\vert^{2}.
\end{equation}
Choosing $\vert \psi(t=t_{f}/2)\rangle$ gives the state exactly in the middle of the time evolution, where, ideally, the target state $\ket{T}$ should be reached with probability one. The time-dependent population in the target state $\vert T\rangle$ is equal to the fidelity defined in Eq. \eqref{eq:fidelity}. Figure \ref{fig:2} presents the target state population at $t=t_f/2$,  for $t_f$ ranging from $0.1 \, \mu$s to $5 \, \mu$s. We observe that, without further improvements, a rather long preparation time is necessary to guarantee adiabatic evolution for both of our sweeps.

\begin{figure}[tb]
   \centering
    \includegraphics[width=\linewidth]{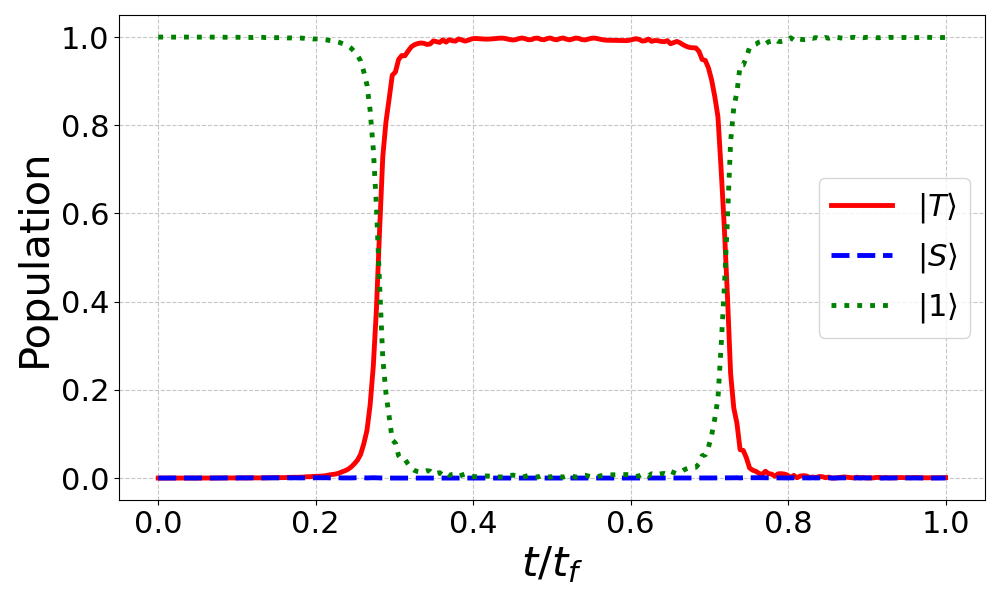}
    \caption{Time dependence of the populations of all  three states for a total protocol time $t_f=5 \, \mu$s. While the overall fidelity is good, corresponding to the plateau in Fig. \ref{fig:2}, tiny oscillations are visible that will only disappear for even longer protocol times.
    }
    \label{fig:3}
\end{figure}

Additionally, we computed the time evolution of the populations of the three states $\vert 1\rangle$, $\vert S\rangle$, and $\vert T\rangle$ for a final time $t_f=5 \, \mu$s, see Fig. \ref{fig:3}. Apart from small fluctuations, the overall evolution is giving acceptable fidelities, comparable for the two sweeping functions in Eqs. \eqref{eq:atan} and \eqref{eq:fit}.

\subsection{Accelerated CD evolution}
\label{sec:CD}

As we have seen, a high-fidelity adiabatic evolution, along the blue level in Fig. \ref{fig:1}, requires a sufficiently long time in order to guarantee a perfect following of the instantaneous eigenstates 
of the time-dependent Hamiltonian $\hat{H}_0(t)$. For practical reasons, those times might be too long, either because of decoherence effects or other experimental problems. This motivates 
one to speed up the evolution, such that the required efficiency is obtained in much shorter times. Ref. \cite{Delvecchio2022} proposed the application of the CD method, which effectively decouples the adiabatic evolution from diabatic transitions to other states than the blue instantaneous level in Fig. \ref{fig:1}.

\begin{figure}[tb]
    \centering
    \includegraphics[width=\linewidth]{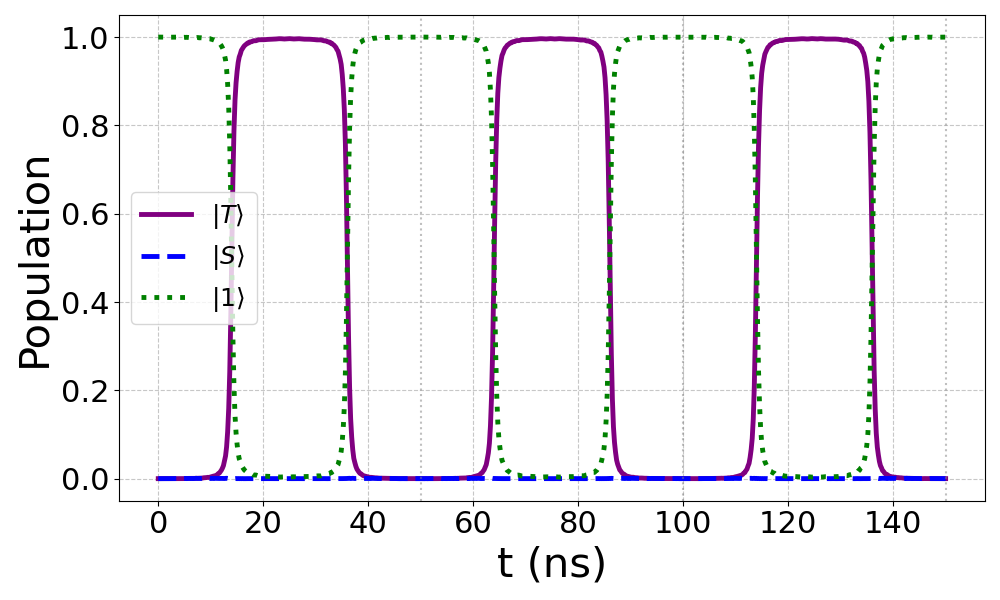}
    \caption{Time evolution with the polynomial sweep $p(\tau)$ and the CD control added. Three repeated cycles are shown. The evolution can now be much faster than in Fig. \ref{fig:3}, here with a single rotation time of $t_{f}=50$ ns. No appreciable loss in the fidelity occurs.}
    \label{fig:4}
\end{figure}

Regarding the three-level model in Eq. \eqref{eq:Ham-1}, we can obtain the CD Hamiltonian $H_{CD}(t)$ by diagonalizing the time-dependent Hamiltonian
\begin{eqnarray}
	\hat H_{0}(t)=\sum_{n} E_{n}(t)\vert n(t)\rangle\langle n(t)\vert \, ,
\end{eqnarray}
where, respectively, $E_{n}(t)$ and $\vert n(t)\rangle$ are the instantaneous eigenvalues and eigenvectors. Then, $H_{CD}(t)$ is given by \cite{Lim1991, Berry2009, Demirplak2003, Demirplak2008, EPL2024}:
\begin{equation}
\label{eq-HCD}
	\hat H_{CD}(t)=i \sum_{n\neq m}\sum_{m}\frac{\vert n(t)\rangle\langle n(t)\vert \partial_{t}\hat H_{0}(t)\vert m(t)\rangle\langle m(t)\vert}{E_{n}(t)-E_{m}(t)} \, .
\end{equation}
The evolution of the system is governed by the total Hamiltonian $\hat H(t)=\hat H_{0}(t)+\hat H_{CD}(t)$, which guarantees that the time evolution adiabatically follows the instantaneous level from which it started (the blue line in Fig. \ref{fig:1} in our case), avoiding transitions to other instantaneous eigenstates. Fig. \ref{fig:4} shows the consequent perfect population transfer from $\vert 1\rangle$ to $\vert T\rangle$ and back, for the polynomial sweep in Eq. \eqref{eq:fit} and a single inversion time of just $t_{f}=50$ ns. This evolution is two orders of magnitude faster than the previous one, for the same parameters used in Fig. \ref{fig:3}.

\subsection{Repeated evolution}
\label{sec-repeat}

The advantage of adding $\hat H_{CD}(t)$ is that the total protocol time can be much faster, while still following the adiabatic state evolution. This allows us to repeat the protocol, that is to rotate many times from the initial state $\ket{1}$ to the state $\ket{T}$ and back again.  The minimal time of a single cycle 
is, in principle, only restricted by the experimental constraints on the pulse  intensities and the rate of the parameter variations,
see \cite{Petiziol2018, Luis2025} for a detailed discussion of the constraints in CD control.
Fig. \ref{fig:4} presents an example of such repeated evolution, for three evolution cycles.  
 The evolution is a hundred times faster than in Fig. \ref{fig:3}, which allows for many cycles of the protocol in an experimentally reasonable time.
 Note that the fidelity in Eq. \eqref{eq:fidelity} is also not noticeably degrading in time, even for a larger number of cycles. In what follows, the CD control is always applied, to guarantee perfect and fast adiabatic evolutions. \edit{The accelerated adiabatic evolution by $H_{CD}$ is less prone to decoherence occurring on longer time scales. The robustness of counterdiabatic schemes under perturbations has been confirmed, see, e.g., \cite{Petiziol2019a, Petiziol2019b, Petiziol2020, Delvecchio2021, Luis2023, Luis2025}.}

\subsection{Adiabatic phase}
\label{sec-phase}

As just shown in the previous section, the evolution $\ket{1} \to \ket{T}$ can be reversed with perfect fidelity. Like in atomic transitions, one may wonder how this reversibility is affected by adding an appropriate phase to the Hamiltonian  in Eq. \eqref{eq:Ham-1}. Following \cite{Barfuss2019}, we explicitly make the matrix element $\Omega_{1T}$ complex. The evolution will then crucially depend on the value and the time dependence of such a phase, see for instance refs. \cite{Hoang2019, Li2021, Kosachiov1992}, with immediate practical consequences discussed, e.g., in \cite{Barfuss2019, Ennio2025}.  For the experiments with molecules described in \cite{Ahmed2011, AhmedLyyra:2013, AhmedLyyra:2014}, the amplitudes $\alpha$ and $\beta$, see appendix \ref{app:A}, quantify the mixing of the singlet and triplet states, due to the Autler-Townes effect \cite{Kirova2005,Ahmed2011, AhmedLyyra:2013, AhmedLyyra:2014}, and can be made complex. This would realize complex coupling elements $\Omega_{1T}=|\Omega_{1T}| \, e^{i \phi}$. \edit{Alternatively, e.g., electro-optical phase shifters could be added to control the coupling matrix elements in a different realization of a three-level system.} In order to engineer the evolution, we make this new phase \edit{of the coupling elements} time-dependent. 

Since in our case we essentially have an evolution in which the third state $\ket{S}$ does not play any role, at least in the ideal adiabatic implementation, we are practically dealing with a two-level system.
\edit{This is similar to the counterdiabatic control of stimulated Raman adiabatic passage (STIRAP) \cite{Petiziol2020, Chen2016, GiannelliArimondo2014}, but here we control two avoided crossings between the same states, not between three different states as in STIRAP.}
We illustrate the resulting evolution in the following, using a Bloch vector model that restricts to the subspace spanned by $\ket{1}$ and $\ket{T}$, and the corresponding effective Hamiltonian
\begin{equation}
\hat H_0^{(\text{eff})}(\tau) =  
\begin{pmatrix}
p(\tau) & |\Omega_{1T}| \,     \mathrm{sign}(\tau - \frac{1}{2}) \, e^{i \phi_k(\tau)} \\
\quad \quad \quad \mathrm{H. \, c.}
& \epsilon_{T} 
\end{pmatrix} \, .
\label{eq:Ham-eff}
\end{equation}
The function $\mathrm{sign}(\tau - \frac{1}{2})$ changes the sign of the coupling along $\sigma_x$ at half the protocol time, while the phase $\phi_k(\tau)$ induces a mixing of the operators $\sigma_x$ and $\sigma_y$. We choose 
\begin{equation}
\label{eq:sign}
    \mathrm{sign}\Big(\tau - \frac{1}{2}\Big) \,  \approx \, \frac{2}{1 + \exp\big[- d \, (\tau - 0.5) \big]} - 1 \,,
\end{equation}
with $d = 20$ for computations without a time-dependent phase $\phi_k(\tau)$. 

\begin{figure}
    \centering
    \includegraphics[width=1\linewidth]{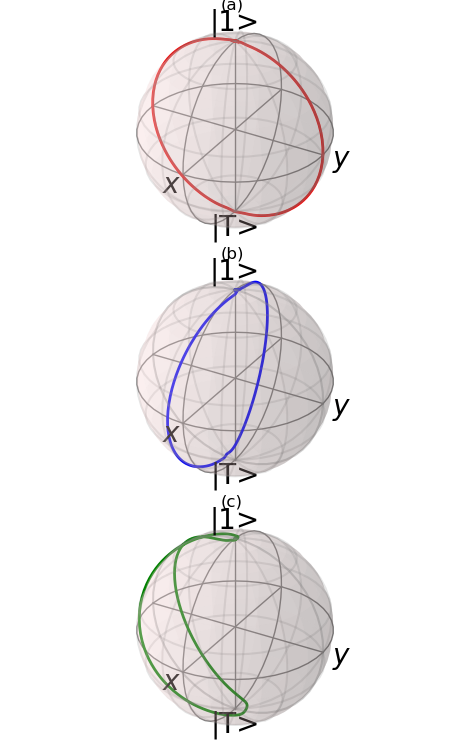}
    \caption{Bloch sphere representation of the dynamics reduced to the subspace spanned by the states $(\ket{1},\ket{T})$. Panel (a): closed loop for vanishing phase $\phi_{k=0} (\tau) = 0$ and a change of sign in the coupling at half the evolution time. Same for a phase evolution with $k=1$ (b) and $k=2$ (c), respectively, without the change of sign.}
    \label{fig:5}
\end{figure}

We can realize a nontrivial closed loop in the evolution on the Bloch sphere with the $\mathrm{sign}$ function and $\phi_k$ constant or zero, or instead without the $\mathrm{sign}$ function \edit{but with} a time-dependent phase $\phi_k(\tau)$. In the latter case, we choose $\phi_k(\tau) = 2 \pi k \tau$ with integer $k$. The corresponding evolutions are illustrated in Fig. \ref{fig:5}, in the panel (a) with $k = 0$ in the presence of the sign change, and in the absence of the $\mathrm{sign}$ change but with two time-dependent choices $k=1$ in (b) and $k=2$ in (c), respectively. The complex phase $\phi_k(\tau)$ effectively then induces a complicated rotation around a time-dependent rotation axis, which lies in the $x-y$ plane of the Bloch sphere.

An evolution along a closed contour on the Bloch sphere, just as shown in Fig. \ref{fig:5}, gives rise to a geometric phase. \edit{Because of the application of the counterdiabatic terms, our evolution is perfectly adiabatic in the instantaneous eigenbasis of the orginal Hamiltonian $H_0$. It is then} interesting to deal with the Berry phase \cite{Berry1984}, arising from the evolution of the diagonal part in the instantaneous eigenbasis $\ket{n(\tau)}$. Berry's phase is expressed in terms of 
  the so-called Berry connection $A(\tau)$, integrated over a closed path drawn by the time evolution. More in detail, $A(\tau)$ is generally defined by the diagonal part of the time evolution in the instantaneous eigenstates $\ket{n(\lambda(t))}$ of $\hat H_{0}(t)$ \cite{EPL2024} when changing some parameter $\lambda(t)$ over time $t$
\begin{equation}
	A(\lambda ) = - i \, %\frac{\partial  \lambda(t)}{\partial t} \, 
    \langle{n (\lambda)} \ket{\frac{\partial n(\lambda)}{\partial \lambda} } \, .
\label{eq-Berry-0}
\end{equation}
Integrating the Berry connection over a closed loop $C$ in the parameter space, we obtain the Berry phase \cite{Berry1984,Berry1987,Zak1989,Wilczek1989}:
\begin{eqnarray}
	\gamma_{\rm B} (T) & = \oint_{C} A(\lambda) \ d\lambda = -i \int_{t=0}^{T} \langle{n(t)} \ket{\frac{\partial n(t)}{\partial t}} \ dt = \nonumber \\
    & = - i\int_{\tau=0}^1 \langle n(\tau) | \frac{\partial n(\tau)}{\partial\tau} \rangle \ d\tau,
	\label{eq-Berry-2}
\end{eqnarray}
where  $t=T=t_f$ is the return time to the initial state  $\ket{1}$.

In the reduced two-state (spin 1/2) model, the absolute value of Berry's phase corresponds to half of the solid angle $\Omega$ enclosed by the motion of the Bloch vector on the Bloch sphere, see e.g. \cite{Griffiths1995, sakurai}. This allows \edit{for} a simple geometric interpretation of Berry's phase. The time-dependent coupling element in Eq. \eqref{eq:Ham-eff} or \eqref{eq:Ham-1}, respectively, is our second time-dependent parameter, next to the sweep function. With these two parameters we indeed can construct evolutions with non-zero solid angle and hence non-vanishing Berry phases \cite{Griffiths1995, sakurai}.

The Berry phases calculated for the two-state model in Eq. \eqref{eq:Ham-eff} coincide with those for the original three-state problem in Eq. \eqref{eq:Ham-1}. For instance, the phase values obtained for the evolution represented in Fig. \ref{fig:5} (b) and (c) are $\gamma_{\rm B}= 2.8$ and $5.5$, respectively. These values reflect the solid angles covered by the closed curves in Fig. \ref{fig:5} (b, blue line) and (c, green line). \edit{The correspondence between the quasi-analytically computed phases using the geometry of the Bloch sphere and the numerically obtained vales for the full three-level system confirms the validity of adiabatic-phase concept. The Berry phase values are, however, }sensitive to parameters, e.g., to the precise form of the sweep function or of the phase $\phi_k(\tau)$. This sensitivity is exploited further in the following subsection.

\subsection{Correlation between Berry phase and state population}
\label{sec:correlation}

The Berry phase sensitively changes when a constant parameter $v$ is added to the polynomial in Eqs. \eqref{eq:fit} and \eqref{eq:Ham-eff}: $p(\tau) \to p(\tau)+v$. The values of $v$ are taken from the interval $[-10,-0.25]$.
The constant shift of the instantaneous energy of state $\ket{1}$ affects the coupling between the instantaneous eigenstates of $\hat H_0^{(\text{eff})}(\tau)$ (and of $\hat H_0 (\tau)$ in Eq. \eqref{eq:Ham-1}\big) during the evolution. This change,  particularly relevant at the avoided level crossings, is visible in the time-dependent fidelity \eqref{eq:fidelity}, as well as in the values of Berry's phase gathered along a cycle.

\begin{figure}[tb]
    \centering
    \includegraphics[width=1\linewidth]{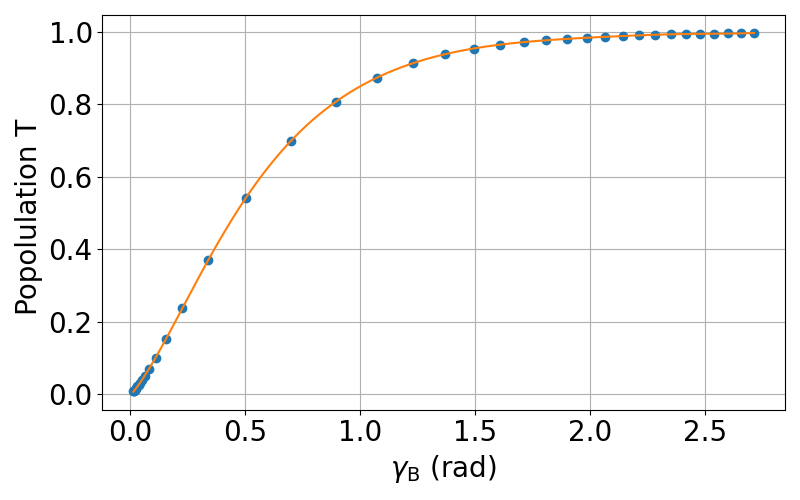}
    \caption{Correlation between the maximal population of state $\ket{T}$ (time-averaged over 20 oscillating maxima when the transfer to the triplet state is far from perfect) and the Berry phase gathered on a cycle, for $k = 1$  in Eq. \eqref{eq:Ham-eff} and slight shifts of the sweep function (and hence of the instantaneous levels) in Eq. \eqref{eq:fit-2} by a constant $v$.  Each red dot corresponds to a different value of $v$ in the interval $[-10 , 0]$ (divided into 41 equidistant steps). The data is fitted by the polynomial function (solid blue line) defined in Eq. \eqref{eq:fit-2}. The transfer protocol starts to break down for phases $\lesssim 2$ giving populations of state $\ket{T}$ visibly smaller than one. The reason for the breakdown is not diabaticity but simply a too small coupling between the states. The coupling strength is governed by the sweep and hence is very sensitive to $v$.}
    \label{fig:6}
\end{figure}

Through this adjustment, it is possible to establish a relationship between the maximal values of the $\ket{T}$ population and the Berry phase. For $v\lesssim -3.5$, the transfer to state $\ket{T}$ 
does not work perfectly anymore. Here we obtain populations of $\ket{T}$ smaller than one, even in the presence of CD control, and hence at perfect adiabatic evolution. The reason is the insufficient coupling between levels $\ket{1}$ and $\ket{T}$. 

The correlation in Fig. \ref{fig:6} is best fitted by a polynomial similar to the applied sweep in Eq. \eqref{eq:fit}
\begin{equation}
\begin{aligned}
P(\gamma_B) ={} & b_8 \gamma_B^8 + b_7 \gamma_B^7 + b_6 \gamma_B^6 + b_5 \gamma_B^5 + b_4 \gamma_B^4 + b_3 \gamma_B^3 \\
                & + b_2 \gamma_B^2 + b_1 \gamma_B + b_0 \, ,
\end{aligned}
\label{eq:fit-2}
\end{equation}
where the coefficients are listed in appendix \ref{app:C}. This relates the maximal population of the triplet state, $P_{T} (t=t_f/2)$ with the Berry phase $\gamma_{\rm B}(T)$ after a full round trip from $t=0$ to $t=T=t_f$. We have chosen this polynomial fit since it has the same form as Eq. \eqref{eq:fit}. The observed correlation indeed seems to reflect the functional form of our sweep $p(\tau)$.

Summing up, the value of the Berry phase can be tuned with the rate of change of the additional phase $\phi_k(\tau)$ (it systematically increases with the integer $k$) and also by the sweep function, for instance, by the constant parameter $v$ added to Eq. \eqref{eq:fit}.

Notice finally that, as visible in Fig. \ref{fig:6}, if the protocol is not perfectly transferring population to the triplet state $\ket{T}$, the geometric phase approaches zero. In the limit of no evolution of state $\ket{1}$, this means that only a vanishing solid angle is spanned in the Bloch spheres as presented in Fig. \ref{fig:5}.

\section{Summary and Outlook}
\label{sec:sum}

The goal of the present work is a reversible, precise, and fast preparation of quantum states that are hard to control with direct transitions \cite{Ahmed2011, Delvecchio2022}. 
We have extended the adiabatic control methods developed by some of us in ref~\cite{Delvecchio2022} to a cyclic transfer of population between an initial (quoted "1") and a final ("triplet" or "T") state, and back to the initial state with high fidelity and short preparation times. The calculations based on the counterdiabatic (CD) protocol drastically improve the protocol with respect to adiabatic sweeps. Results for repeated fast evolution over a few cycles with high fidelity are presented. This shows that repetitions of the CD protocol speed up by a hundred times without loss of fidelity are possible. 

In addition, we investigated a topological (Berry's) phase, arising from the evolution along the closed contour on the Bloch sphere, and established a relationship between the maximal population of the triplet state and the Berry phase. Reciprocity is controlled by the sweep function and an additional time-dependent complex phase in the Hamiltonian. 

Our protocols are relevant for future experimental applications in quantum optics and quantum technologies. Some examples include transfer between singlet and triplet populations in the process of cold molecule formation \cite{B2024}, or the implementation and improvement of an all-optical spin switch, initially proposed in \cite{AhmedLyyra:2014} and realized in \cite{StahovichAhmed2024} with different methods. \edit{Extensions of our work to more complex quantum systems are possible as long as the avoided crossings between the relevant states are well separated in time and can hence be controlled independently.}

%\medskip

\section*{Acknowledgements}
We thank Michele Delvecchio, Michele Burrello and Simon Dengis for useful discussions. L.L. thanks Stefano Carretta and Paolo Santini for support. E.A., T.K., and S.W. acknowledge funding from Q-DYNAMO (EU HORIZON-MSCA-2022-SE-01) with project No. 101131418, and L.L. and S.W. from the National Recovery and Resilience Plan through Mission 4 Component 2 Investment 1.3, Call for tender No. 341 of 15/3/2022 of Italian MUR funded by NextGenerationEU, with project No. PE0000023, Concession Decree No. 1564 of 11/10/2022 adopted by MUR, CUP D93C22000940001, Project title "National Quantum Science and Technology Institute" (NQSTI).

\appendix

\section{The experimental scheme}
\label{app:A}

The singlet-triplet transfer experiment described in \cite{Ahmed2011,AhmedLyyra:2013,AhmedLyyra:2014}, reduces to the four-level scheme introduced in \cite{Kirova2005} and plotted schematically in Fig. 1 of \cite{Delvecchio2022}. The original level structure for a lithium dimer is shown, e.g., in Fig. 2 in \cite{AhmedLyyra:2013}. Our initial ground state $|1\rangle$ would correspond in Li$_2$ to the rotational-vibrational state $X^1 \Sigma_g^+(\nu= 2, J = 22)$, and the spin-orbit coupled states $\ket{T}$ and $\ket{S}$ to $1^3 \Sigma_g^-(\nu= 1, J= 21, f)$ and $G^1\Pi_g(\nu=12,J=21,f)$, respectively. In the experiment \cite{AhmedLyyra:2013}, the first transition from the ground to the excited state is a two-photon one. The goal of \cite{AhmedLyyra:2013, Delvecchio2022} is to transfer the population from the singlet initial state $|1\rangle$ to the triplet excited state $|T\rangle$. What helps here is that the $(|S\rangle,|T\rangle)$ manifold of states obeys  mixed singlet-triplet symmetry owing to spin-orbit coupling. They originate from the $(|S_0\rangle,|T_0\rangle)$  states corresponding to unperturbed singlet and triplet states with original energy separation $\Delta_0$. These states experience a spin-orbit perturbation with amplitude $V$ described by the Hamiltonian $H_{\rm so}= V|S_0\rangle\langle T_0|+ H.c$, (assuming $\hbar=1$). The mixed states $|S\rangle$ and $|T\rangle$ are then given by
\begin{equation}
\label{eq:STeigens}
|S\rangle =\alpha |S_0\rangle- \beta |T_0\rangle, \qquad
|T\rangle =\beta |S_0\rangle+\alpha |T_0\rangle ,
\end{equation}
where the $(\alpha,\beta)$ coefficients are normalized to one, \textit{i.e.}, $|\alpha|^2+|\beta|^2=1$. The unperturbed energy separation $\Delta_0$ and the perturbation $V$ are linked to the effective energy splitting $\Delta_{so}$ of the $(|S\rangle,|T\rangle)$ levels and to the mixing coefficients by \cite{Kirova2005}
\begin{equation}
\label{eq:spinorbit}
\Delta_0= (\alpha^2-\beta^2)\Delta_{so} \qquad
V = \alpha\beta\Delta_{so}.
\end{equation}
The Lithium molecular states of \cite{Ahmed2011} have, for instance, mixing coefficients $\alpha^2=0.87$ and $\beta^2=0.13$, with spin-orbit splitting $\Delta_{so} = 2\pi\cdot 0.75$ GHz equivalent to $4.71$  ns$^{-1}$. 

Access to the singlet-triplet manifold is provided by a pump ($p$) laser connecting the $\ket{1}$ singlet to the $|S\rangle$ state with detuning $\delta_p$.  The singlet component of both  $|S\rangle, |T\rangle$ eigenstates of Eq.~\eqref{eq:STeigens} determines their excitation by the $p$ laser characterized by Rabi frequencies $\alpha\Omega_p$ and $\beta\Omega_p$, respectively. The effective excitation is controlled by the pump detuning from those states. 

The  quantum control of the singlet-triplet transfer introduced in \cite{Ahmed2011,AhmedLyyra:2013} is based on the modification of the energy separation of the spin-orbit $|S\rangle, |T\rangle$ manifold. Such a modification is produced by a control  ($c$) off-resonant laser linking the $|T\rangle$ state to an additional $|2\rangle$ triplet one with detuning $\delta_c$.  The Rabi frequencies for the mixed manifold components are $-\beta\Omega_c$ and $\alpha\Omega_c$, respectively.

Overall, the temporal evolution of the optically driven four levels $(|1\rangle,|S\rangle,|T\rangle, |2\rangle)$ is  described by the Hamiltonian
 \begin{equation} 
\label{eq:Hmatrix4level} 
H^{(4)}=
 \begin{pmatrix}
    \delta_p& \alpha\Omega_p/2 &-\beta\Omega_p/2&0\\
    \alpha\Omega_p/2 & 0 &0& \beta \Omega_c/2\\
    -\beta\Omega_p/2 & 0 &-\Delta_{so}&\alpha \Omega_c/2 \\
    0 & \beta\Omega_c/2 &\alpha \Omega_c/2  &-\delta_{c}-\Delta_{so}
  \end{pmatrix}.
\end{equation}
One assumes that the zero energy corresponds to the $|S\rangle$ state and defines $\delta_p=\omega_p-(E_S-E_1)$, $\delta_c=\omega_c-(E_2-E_T)$, with $E_i$ the energy of the states $(i=1,2,S,T)$. In the experiment \cite{Ahmed2011}, the Rabi frequencies were, e.g., $\Omega_p=0.24$ and $\Omega_c=3.8$ in units of ns$^{-1}$.

\subsection{The state reduction from four to three levels}
\label{app:A-1}

As shown in \cite{Delvecchio2022}, for  $|\delta_{c}|\gg \Omega_c$, corresponding to a virtual excitation of the $|2\rangle$ state, the $\ket{2}$ state can be adiabatically eliminated. The corresponding reduced three-level Hamiltonian is
\begin{equation} 
\label{eq:Hmatrix3level} 
H^{(3)}=
  \begin{pmatrix}
    \delta_p &\frac{\alpha\Omega_p}{2} & -\frac{\beta\Omega_p}{2}\\
    \frac{\alpha\Omega_p}{2} & \frac{\beta^2\Omega_c^2}{4(\delta_{c}+\Delta_{so})} &\frac{\alpha\beta\Omega_c^2}{4(\delta_{c}+\Delta_{so})} \\
    -\frac{\beta\Omega_p}{2} & \frac{\alpha\beta\Omega_c^2}{4(\delta_{c}+\Delta_{so})} &-\Delta_{so} +\frac{\alpha^2\Omega_c^2}{4(\delta_{c}+\Delta_{so})}\\
  \end{pmatrix}\,,
\end{equation}
now in the basis $\{\ket{1},\ket{S},\ket{T}\}$. 
Notice that within the effective energy of the state $|T\rangle$ the light-shift or ac shift $\delta_{ls}$ given by 
\begin{equation}
\label{eq:lightshift}
\delta_{ls}=\frac{\alpha^2\Omega_c^2}{4(\delta_{c}+\Delta_{so})},
\end{equation} 
associated to the Autler-Townes process.  

Please note that, in contrast to \cite{Delvecchio2022}, above the
transformation to the rotating frame oscillating with frequency $\Delta_{so}$ was done before the adiabatic elimination of the fourth state. This approximation is then more accurate than doing it the other way around, in which the matrices are \edit{originally} given in \cite{Delvecchio2022}.

\subsection{Matrix elements}
\label{app:A-2}

The matrix elements of Eq. \eqref{eq:Ham-1} relate to the parameters \edit{given in Eq. \eqref{eq:Hmatrix3level}} from \cite{Delvecchio2022}, in the following manner:

\begin{eqnarray}
    \label{app:1}
    \epsilon_{1}(t)=\delta_{p}(t) \, ,\,
    \epsilon_{S}=\frac{\beta^{2} \, \Omega_{c}^{2}}{4(\delta_{c}+\Delta_{so})} \, , \\
    \epsilon_{T}=-\Delta_{so}+\frac{\alpha^{2} \, \Omega_{c}^{2}}{4(\delta_{c}+\Delta_{so})} \, , \\
    \Omega_{1S}=\frac{\alpha \, \Omega_{p}}{2} \, , \,
    \Omega_{1T}=-\frac{\beta \, \Omega_{p}}{2} \, , \\
    \Omega_{ST}=\frac{\alpha \, \beta \, \Omega_{c}^{2}}{4(\delta_{c}+\Delta_{so})} \, .
\end{eqnarray}

\section{Polynomial sweep function}
\label{app:B}

 The coefficients of the polynomial fit of Eq. \eqref{eq:fit} to Eq. \eqref{eq:atan} are given by
 
%\begin{align}
  %      & a_{8}=-5.79443 \times 10^4 ,\,  a_{7}=2.31777 \times 10^5,\\
   %     & a_{6}=-3.93182 \times 10^5 ,\, a_{5}=3.68326 \times 10^5, \\
    %    & a_{4}=- 2.07989 \times 10^5 ,\, a_{3}=7.25080 \times 10^4, \\
     %   & a_{2}=- 1.54206 \times 10^4 ,\, a_{1}=1.92473 \times 10^3, \\
      %  & a_{0}=- 1.55700 \times 10^2.
 %   \end{align}
\begin{align}
        & a_{8} = -6.95332 \times 10^3 ,\quad  a_{7} = 2.78132 \times 10^4,\\
        & a_{6} = -4.71818 \times 10^4 ,\quad a_{5} = 4.41991 \times 10^4, \\
        & a_{4} = -2.49587 \times 10^4 ,\quad a_{3} = 8.70096 \times 10^3, \\
        & a_{2} = -1.85047 \times 10^3 ,\quad a_{1} = 2.30968 \times 10^2, \\
        & a_{0} = -1.86840 \times 10^1.
\end{align}

The parameter $v$ in Sec. \ref{sec:correlation} is scanned in the interval $[-10 , -0.25]$ in 40 equidistant steps.

\section{Polynomial fit to correlation}
\label{app:C}

The coefficients of the polynomial fit, Eq. \eqref{eq:fit-2}, in the relation between the maximal populations of
the state $\ket{T}$ and the values of the Berry phase are given by

\begin{align*}
b_0 &= -0.00435 \hspace{1cm} & b_1 &= 0.74855 \\
b_2 &= 2.41404 \hspace{1cm} & b_3 &= -5.60260 \\
b_4 &= 5.62913 \hspace{1cm} & b_5 &= -3.26138 \\
b_6 &= 1.12098 \hspace{1cm} & b_7 &= -0.21196 \\
b_8 &= 0.01696 .
\end{align*}

%%%%%%%%%%%%%%%%%%%%%%%%%%%%%%%%%%%%%%%%%%%%%%%%%%%%%%

%\bibliography{ref}

\end{document}